# How Retrocausality Helps

Roderick I. Sutherland

Centre for Time, University of Sydney, NSW 2006 Australia

rod.sutherland@sydney.edu.au

**Abstract**

It has become increasingly apparent that a number of perplexing issues associated with the interpretation of quantum mechanics are more easily resolved once the notion of retrocausality is introduced. The aim here is to list and discuss various examples where a clear explanation has become available via this approach. In so doing, the intention is to highlight that this direction of research deserves more attention than it presently receives.

**Introduction**

While quantum mechanics is a highly successful mathematical theory in terms of experimental verification, there remain long-standing questions as to what sort of physical reality could be underlying the mathematics and giving rise to the theory's stranger predictions. Unlike classical mechanics, the theory does not give sufficient guidance towards identifying the appropriate ontology. Over time, there has been a growing awareness that backwards-in-time influences, or retrocausality, might be relevant in interpreting and understanding some of the phenomena in question. The intention here is to summarise some of the advantages that can be gained by introducing retrocausality into the underlying picture. In particular, it is found that taking this step can achieve the following:

1. It can restore locality in the case of entangled states (such as with Bell's theorem)
2. It can preserve consistency with special relativity at the ontological level
3. It can allow replacement of many-particle, configuration space wavefunctions by individual wavefunctions
4. In can allow statistical descriptions to be replaced by definite, ontological values
5. It can facilitate the development of a Lagrangian formulation in the case where a particle ontology is assumed
6. It can suggest significant improvements to existing ontological models.

These points will be discussed individually in the following sections. A first step, however, is to define more precisely what is meant by retrocausality here. In doing so, it should be noted that no suggestion is being made of movement through 4-dimensional spacetime in either the forwards or backwards time directions. Motion remains confined, as usual, to the 3-dimensional picture. In this context, the definition of retrocausality will be taken to be as follows:

> It is necessary to specify final boundary conditions as well as the usual initial ones in order to determine the state completely at any intermediate time, with the experimenter's controllable choice of the final conditions thereby exerting a backwards-in-time influence.

For further clarity, two conventions will be introduced at this point:

(i) Since any initial boundary condition in standard quantum mechanics is specified by a Hilbert space vector $|i\rangle$, it will be presumed by symmetry that any final boundary condition should be similarly specified by a Hilbert space vector $|f\rangle$.

(ii) The initial boundary condition here will simply be equated with the result of the most recent prior measurement performed on the system. Similarly, any final boundary condition will be equated with the result of the next measurement performed. (This is not intended to imply any special status for measurement interactions compared with other interactions, but merely to frame the discussion in a clear-cut form.)

These two conventions keep the mathematics simple and straightforward. Using the second one, the definition of retrocausality can be formulated more specifically as:

- The choice of observable measured at a particular time can affect the hidden ontology existing at an earlier time.

Having defined retrocausality in this way, various advantages it provides will now be discussed.

**1. Locality can be restored in the case of entangled states**

This is the best-known case in which retrocausality is suspected to be relevant and so need only be outlined briefly here. It will be illustrated via the well-known arrangement employed in Bell's theorem [1]. This involves two particles that are in an entangled state due to a previous interaction or decay process, but have now ceased interacting and moved far apart. This situation is represented on the spacetime diagram in Fig. 1. Under certain reasonable assumptions, the theorem implies that the choice of measurement at the point labelled $M_1$ on the diagram must have an effect on the result of the measurement performed at $M_2$. This apparently nonlocal connection between the well-separated events at $M_1$ and $M_2$ is perplexing at first sight, but can be given a local explanation once retrocausality is permitted [2]. Specifically, the effect of the measurement choice at $M_1$ is taken to be communicated along the path $M_1DM_2$ on the diagram. The apparent nonlocality in 3 dimensions then becomes local from a 4-dimensional viewpoint as a result of allowing a backwards-in-time link along the path $M_1D$. Although this explanation has been formulated for the particular case of Bell's theorem, it can be applied to any entangled state.

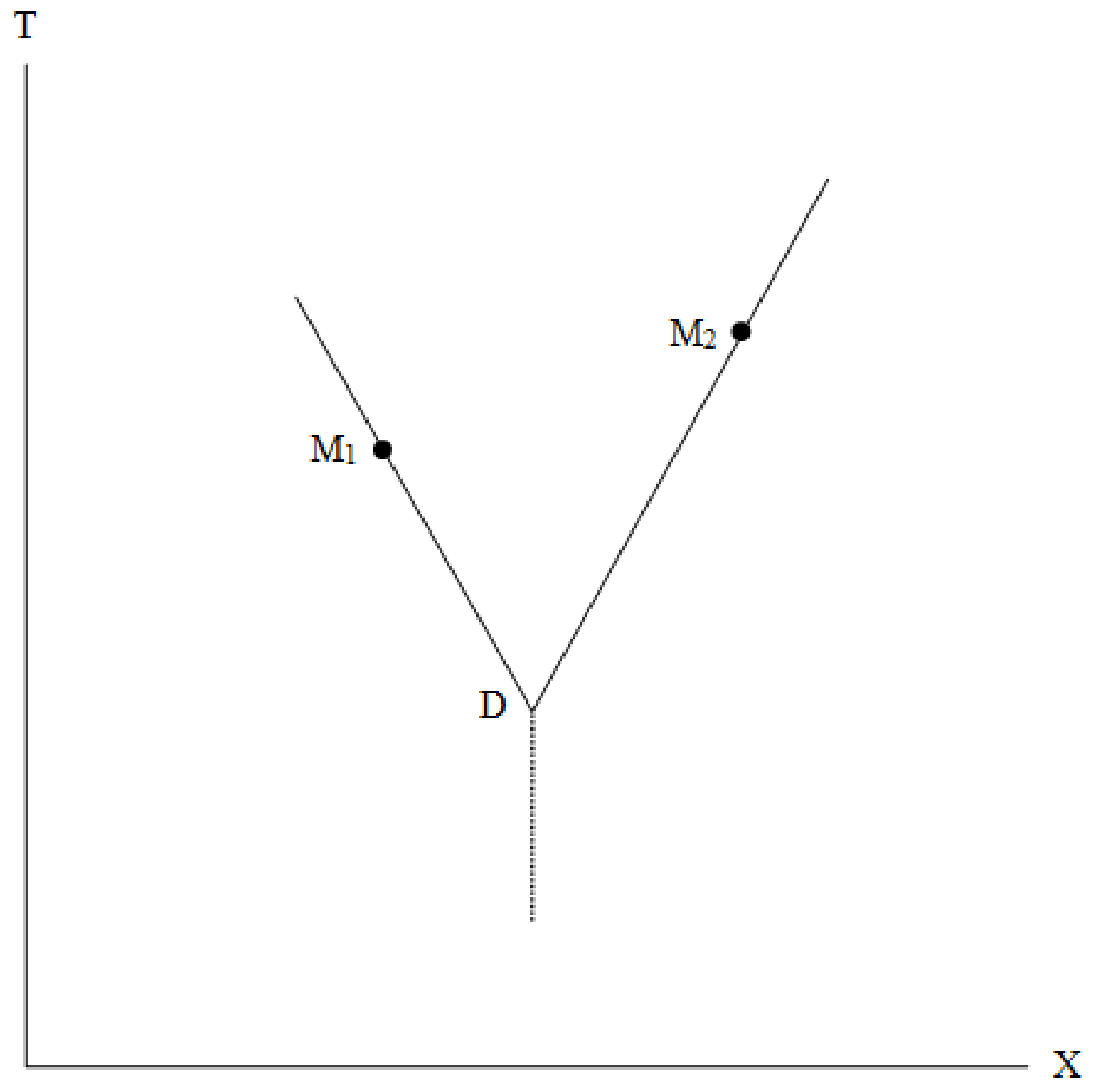

Fig. 1

**2. Consistency with special relativity can be preserved**

The question of consistency with relativity is closely related to the previous point and will again be discussed in terms of the set-up for Bell's theorem. The apparent non-local communication seemingly implied by that theorem suggests the need for a spacelike signal passing between the particles. This would run contrary to special relativity because it would require identifying a particular spacetime hyperplane along which the communication propagates, thereby singling out a preferred reference frame. An example of a model where such a preferred frame must be incorporated is Bohmian mechanics. In contrast, the zig zag path invoked in Fig. 1 is clearly Lorentz invariant. Hence introducing retrocausality by allowing this path not only restores locality but also maintains special relativity at the ontological level.

In similar fashion, Lorentz invariance can be preserved when considering the apparently instantaneous reduction of a two-particle wavefunction to individual single-particle wavefunctions as a result of a measurement (e.g., by one of the measurements in Fig. 1). This reduction again seems to require a preferred frame, but the matter can be resolved once more with the aid of retrocausality, as will be discussed in more detail in the next section.

## 3. Entangled configuration-space wavefunctions can be replaced by individual wavefunctions

A technique for replacing many-particle wavefunctions with individual wavefunctions (before any measurements actually produce such a factorisation) is outlined here and could be useful in constructing a range of different models in relation to the interpretation quantum mechanics. To understand what is involved, consider the following description provided by standard quantum mechanics. Suppose two particles are initially in an entangled state $\psi(\mathbf{x},\mathbf{x}';t)$ at time t, where the position coordinates $\mathbf{x}$ and $\mathbf{x}'$ refer to the 1st and 2nd particle, respectively. Any measurement performed on the 1st particle will then yield a separate eigenfunction $\psi_1(\mathbf{x};t)$ for this particle. The 2nd particle will also acquire a separate wavefunction as a result of this measurement. This updated state of the 2nd particle will be given by[1]:

$$\psi_2(\mathbf{x}';t) = \frac{1}{N}\int \psi_1(\mathbf{x};t)\,\psi(\mathbf{x},\mathbf{x}';t)\,d^3x$$

where N is a normalisation constant. (This useful equation is not usually stated in text books.) Now the standard theory assumes that this state will arise simultaneously with the measurement on the 1st particle. But "simultaneous" is ambiguous here in a relativistic context - it would require a preferred reference frame. Also, in some other frames, such a collapse at a spacelike location would be viewed as occurring **before** the measurement.

To avoid the above difficulty, a Lorentz invariant description would be preferable. Surprisingly, such a description becomes easily available[2] once retrocausality is accepted into the picture. For ontological purposes, one can simply assume that the 2nd particle's updated wavefunction is already applicable from the time when the two particles originally separated. This simple assumption has no obvious negative consequences and provides three immediate benefits:

(i) the configuration space wavefunction is avoided completely

(ii) locality is maintained because there is no wavefunction collapse at a distant point

(iii) special relativity is respected, since there is no preferred frame.

Note, however, that this trick is only permissible if retrocausality is allowed, since the form of the new wavefunction depends on a **later** choice of measurement on the other particle. In the general case of n entangled particles, an individual wavefunction can successfully be assigned to each particle in this way, but each wavefunction will depend on the later measurements performed on all the other particles.

---

[1] The non-relativistic expression is used here for simplicity. The relativistic case is given in Ref. [3], Eq. (2).
[2] ibid, Sec. 3.

As is shown elsewhere[3], this separability technique is quite consistent with the usual statistical correlations predicted by quantum mechanics. The basic point is that the required correlations reappear once the unknown future influences are averaged out. This technique gives a simple and general procedure for transforming from configuration space to spacetime in any case where particles states are entangled from previous interactions. The method is independent of whatever underlying ontology one prefers (e.g., particles or fields) and could be a useful first step in developing further models. The key point, however, is that it would not be available at all without retrocausality.

**4. Some statistical descriptions can be replaced by definite, ontological values**

This point will be illustrated by taking the energy-momentum tensor $T^{\mu\nu}$ as an example, this tensor being of importance in attempts to formulate a theory of quantum gravity. Generally, in the transition from classical to quantum mechanics, the description of observable quantities necessarily becomes statistical. The energy-momentum tensor is no exception, as can be seen from the fact that it becomes dependent on the wavefunction $\psi(x)$:

$$T^{\mu\nu}(x) = \psi^*(x)\hat{T}^{\mu\nu}\,\psi(x)$$

with this expression subject to abrupt change upon measurement (here $\hat{T}^{\mu\nu}$ is the energy-momentum operator). Looking at the Einstein field equation for gravity:

$$G^{\mu\nu} = 8\pi T^{\mu\nu}$$

which relates $T^{\mu\nu}$ to the spacetime curvature described by the Einstein tensor $G^{\mu\nu}$, the fact that the right hand side has become a statistical quantity requires that $G^{\mu\nu}$ on the left hand side must also become quantised and statistical to maintain consistency.

There is, however, a retrocausal alternative. Switching to Dirac notation, one can choose the energy-momentum tensor to be defined by the following equation [4]:

$$T^{\mu\nu}(x) = \mathrm{Re}\frac{\langle f|x\rangle \hat{T}^{\mu\nu}\langle x|i\rangle}{\langle f|i\rangle}$$

This expression then provides the energy and momentum density at any point x given both the initial and final boundary conditions i and f. It is similar to the "weak value" expressions of Aharonov, Albert and Vaidman [5], although the real value has been taken to be consistent with real spacetime. It reduces back to the standard expression for $T^{\mu\nu}$ once a weighted average is taken over the unknown final state f. The key point to note is that the new expression can be viewed as a definite, ontological quantity. It does not merely provide a statistical description for the outcome of the next measurement - it already contains the actual outcome f and so the probabilities of other outcomes are irrelevant. Since $T^{\mu\nu}$ is now definite, this means that $G^{\mu\nu}$ on the other side of the Einstein equation no longer needs to be statistical. Pursuing this approach, the resulting theory can yield definite (rather than “fuzzy”) values for the curvature. It also avoids the basic problem of trying to get started on building a quantum gravity model without having a pre-existing spacetime background.

---

[3] ibid, Sec. 8.

Whether this approach is on the right track or not, the essential point is that such a simple, alternative theory would not be available without retrocausality.

**5. A Lagrangian formulation can be constructed in the case of particle ontologies**

For models in which an underlying ontology of particle trajectories is proposed, retrocausality enables a Lagrangian description to be formulated in 4-dimensional spacetime, even for the many-particle case [3]. This involves employing some of the methods already described in the previous sections. It has the immediate advantage that one can easily derive the relevant field equations, particle equations of motion, conservation laws, energy-momentum tensors, currents, etc., all from a single scalar expression. This convenient and more comprehensive mathematical formalism can then answer any question one wishes to ask. The formalism is also easily set in Lorentz invariant form, which would not be possible without retrocausality.

**6. Improvements to existing models become apparent**

As an example of this possibility, the well-known Bohm model will be considered. This model assumes that particles have definite trajectories at all times and are guided by an accompanying field. Some possible weaknesses that could be claimed for this model are that it is not easily generalised to relativistic cases, that there is no apparent source for the guiding field and that energy and momentum are not conserved. Also, in the many-particle case, the model has the possible deficiencies that it is necessarily nonlocal, that it requires a preferred reference frame and that reality seems to reside in 3n-dimensional configuration space.

In response to these points, a retrocausal version of Bohm’s model has been formulated [6,3], thereby providing the following possible improvements in the ontology:

1. The model is Lorentz invariant

2. A general form applicable for any wave equation is possible

3. The model is local from a spacetime viewpoint

4. Energy and momentum conservation are restored

5. Reality resides in 4-dimensional spacetime instead of configuration space

6. The correct statistical correlations can be maintained while employing a separate wavefunction for each particle

7. Each particle in an entangled state has a separate velocity expression, instead of just a single, overall 3n-dimensional velocity

8. A physical interpretation can be provided [7] for the negative values of the Klein-Gordon "probability" density

9. The guiding field can be given a possible source (viz., the associated particle).

It is, of course, a matter of taste which version (if any) of the Bohm model is preferred, but the essential point again is that such a choice only exists as a result of contemplating retrocausality. Indeed, this short paper has endeavoured to make clear that, regardless of one's preferred ontological picture for quantum mechanics, retrocausality introduces a number of promising possibilities which were previously unavailable.